\documentclass[10pt,aip,twocolumn]{revtex4-1}
\newcommand\p{\partial } 
\usepackage{fullpage}
\usepackage{amssymb}
\usepackage{amsmath}
\usepackage{hyperref}
\usepackage{natbib}
\usepackage{color}
\usepackage{graphicx}
\usepackage{graphics}

\begin{document}

\title{Characterizing Fluid and Kinetic Instabilities using
  Field-Particle Correlations on Single-Point Time Series}
\author{Kristopher ~G. Klein} \email{kriskl@umich.edu}
\affiliation{Department of Climate and Space Sciences and Engineering,
  University of Michigan, Ann Arbor, MI 48109, USA}


\begin{abstract}
A recently proposed technique correlating electric fields and particle
velocity distributions is applied to single-point time series
extracted from linearly unstable, electrostatic numerical
simulations. The form of the correlation, which measures the transfer
of phase-space energy density between the electric field and plasma
distributions and had previously been applied to damped electrostatic
systems, is modified to include the effects of drifting equilibrium
distributions of the type that drive counter-streaming and
bump-on-tail instabilities. By using single-point time series, the
correlation is ideal for diagnosing dynamics in systems where access
to integrated quantities, such as energy, is observationally
infeasible.  The velocity-space structure of the field-particle
correlation is shown to characterize the underlying physical
mechanisms driving unstable systems.  The use of this correlation in
simple systems will assist in its eventual application to turbulent,
magnetized plasmas, with the ultimate goal of characterizing the
nature of mechanisms that damp turbulent fluctuations in the solar
wind.
\end{abstract}

\maketitle

\section{Introduction}
\label{sec:intro}
A significant goal of plasma physics research is the characterization
of mass, momentum, and energy transport in a wide variety of complex
systems. In particular, the question of what mechanisms mediate the
transfer of energy between turbulent fields and distributions of
plasma particles, leading to the eventual damping and dissipation of
turbulence, is open. One system which displays turbulent behavior is
the solar wind, a hot, diffuse emanation from the Sun's surface that
fills the heliosphere. While lacking the precise control over
conditions afforded in a laboratory setting, the large volume of in
situ measurements of the solar wind over the last half century has led
to the accrual of observations of turbulence with a wide variety of
plasma parameters. Such measurements have proven useful to the study
of phenomena in magnetized turbulence.\cite{Bruno:2005}

A limitation of these in situ observations is that they are taken at a
single-point in space at a given time.\footnote{Notable exceptions to
  the single-point limitation are the are the
  \textit{Cluster},\cite{Escoubet:2001}
  \textit{THEMIS},\cite{Angelopoulos:2008} and \textit{Magnetospheric
    Multiscale (MMS)}\cite{Burch:2016} missions, which are comprised
  of four or five spacecraft arranged in particular geometric
  configurations.}  This raises at least two significant
complications; one must disentangle the dynamics associated with
spatial and temporal variation, and track the evolution of spatially
integrated quantities, such as the energy content of a field or
distribution of charged particles, given only single-point
measurements. The first of these complications is addressed by
invoking Taylor's Hypothesis,\cite{Taylor:1938} the conjecture that
for single-point measurements of sufficiently fast flows, the time
evolution is essentially frozen and the measurement traces out the
spatial structure of the turbulence; a review of the application of
Taylor's Hypothesis to solar wind observations can be found in Klein
et al 2014.\cite{Klein:2014b}

To address the second complication of inaccessible spatially
integrated quantities, one may consider the dynamics of the
phase-space energy density rather than the total energy. A technique
has been recently proposed to measure the local-in-phase-space energy
transfer between fields and plasma distributions using single-point
time series of simple plasma systems.\citep{Klein:2016a,Howes:2016} By
correlating the product of the electric field and velocity derivative
of the particle distribution measured at a single point in space, the
velocity structure of the transfer of energy between fields and
particles is obtained. By averaging this correlation over a selected
time interval, the oscillatory energy transfer between the fields and
particles is removed, leaving only the secular energy transfer. The
mechanisms responsible for this energy transfer can be identified by
the velocity space structure of the field-particle correlation.
Initial work applied this correlation to systems that damp via the
Landau resonance.\cite{Landau:1946} Here, we consider the transfer of energy in linearly
unstable systems, and show that field-particle correlations are able
to identify the presence of both fluid and kinetic instabilities in
such systems.

The unstable systems under consideration are reviewed in
Section~\ref{sec:drifts}. In Section~\ref{sec:vp}, we present the
nonlinear numerical code employed in this work, \texttt{VP}, as well
as the three simulations under consideration. In
Section~\ref{sec:fpc}, the field-particle correlation is presented and
applied to the three simulations.  Analysis and discussion related to
the correlations are found in Section~\ref{sec:method}. Application of
field-particle correlations to simple, homogeneous, linearly unstable
systems enables the construction of signatures of basic energy
transfer mechanisms. Combined with signatures for all relevant energy
transfer mechanisms, such correlations may be usefully employed to
diagnose the behavior of more complex systems.

\begin{figure*}[ht]
\begin{center}
\includegraphics[width=16.5cm,viewport=5 5 385 125, clip=true]
{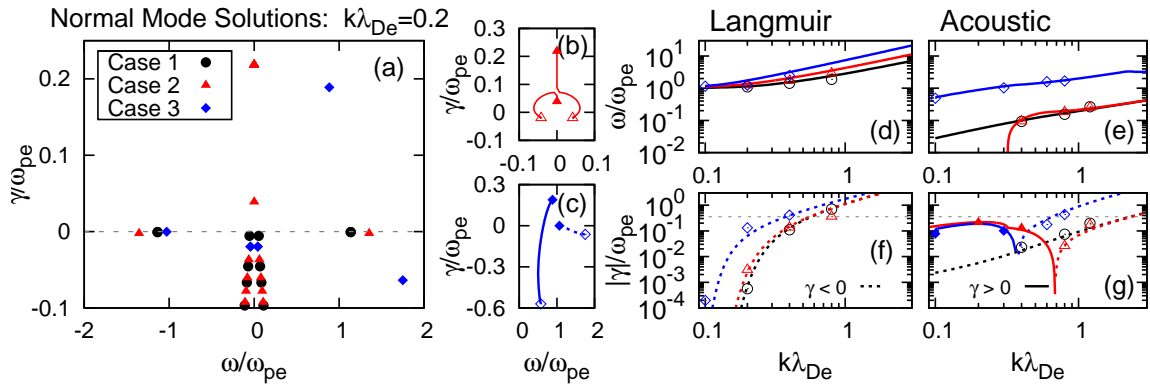}
\caption{Normal mode properties for the three cases examined in this
  work.  The complex frequency solutions to Eqn~\ref{eqn:disp.fpc} for
  $k\lambda_{De}=0.2$ and the parameters of cases 1, 2, and 3 (black,
  red, and blue) are shown in panel a. Parametric paths for normal
  mode solutions from $|v_{dej}|=0$ $(v_{de2}=0)$ to $|v_{dej}|=1.75$
  $(v_{de2}=4.25)$ are plotted in panel b (c), indicating the
  connection between stable and unstable acoustic modes. Frequency
  dispersion curves $\omega(k\lambda_{De})/\omega_{pe}$ for the least
  damped and/or fastest growing Langmuir and acoustic modes are
  illustrated in panels d and e. Damping rates
  $\gamma(k\lambda_{De})/\omega_{pe}$ for the same modes are plotted
  in panels f and g, with solid (dashed) lines indicating growth
  (damping). Frequency estimates extracted from linear and small
  amplitude nonlinear \texttt{VP} simulations are shown as points in
  panels d-g.}
\label{fig:linear}
\end{center}
\end{figure*}

\section{Linearly Unstable Systems}
\label{sec:drifts}
The 1D1V electrostatic systems of interest in this work are governed
by the Vlasov and Poisson equations
\begin{equation}
\frac{\p f_s}{\p t} + 
v \frac{\p f_s}{\p x} -
\frac{q_s}{m_s}\frac{\p \phi}{\p x} \frac{\p f_s}{\p v}  = 0,
\label{eqn:vlasov.ld}
\end{equation}
and
\begin{equation}
\frac{\p^2 \phi}{\p x^2} =
-4 \pi \sum_s q_s \int_{-\infty}^\infty dv f_s
\label{eqn:poisson.ld}
\end{equation}
which evolve the distribution of species $s$, $f_s(x,v,t)$ and the
electric field $E(x,t)=-\p\phi(x,t)/\p x$.  Recent work has applied
field-particle correlations to electrostatic stable systems as a means
of extracting the velocity dependent signature of Landau damping from
single-point time series.\citep{Klein:2016a,Howes:2016} Here, we
consider electrostatic systems unstable to either fluid or kinetic
instabilities, and characterize the signature of the associated growth
using field-particle correlations. For fluid instabilities, the
behavior of the plasma depends on the bulk parameters of the system
and the energy transfer is not organized by characteristic velocities
such as the resonant wave phase speed; energy transfer for kinetic
instabilities depends on such characteristic velocities, as the
electrostatic field acts to exchange particles of higher and lower
kinetic energies near the field's resonant velocity. For both types of
instabilities, one or more of the distributions lose energy, while the
fields and other distributions gain energy, resulting in an inverse
damping. General discussion of plasma instabilities, as well as
particular treatments of the electron drift instabilities of interest
in this work, can be found in many plasma
textbooks.\cite{Krall:1973,Stix:1992,Hazeltine:2004}

\begin{table}[h]
\caption{Electron Bulk Parameters}
\begin{center}
\begin{tabular}{|c|c|c|c|c|}
\hline
& $n_{e1}/n_{i}$ & $v_{d1}/v_{te}$ & $n_{e2}/n_{i}$ & $v_{d2}/v_{te}$ \\
\hline
Case 1 & $0.5$ & $0.75$ & $0.5$ & $-0.75$ \\ 
Case 2 & $0.5$ & $1.75$ & $0.5$ & $-1.75$ \\ 
Case 3 & $0.9$ & $0.00$ & $0.1$ & $ 4.25$ \\
\hline
\end{tabular}
\end{center}
\label{tb:params}
\end{table}

To highlight the distinct velocity space structure of energy transfer
in unstable systems, we consider three cases, each with distinct sets
of parameters for one population of ions and two populations of
electrons. The equilibrium distributions for the three populations
take the Maxwellian form
\begin{equation}
F_{s0}(v)=\frac{n_{sj}}{\sqrt{2\pi}}\exp\left[ \frac{-\left(
    v-v_{dsj}\right)^2}{2 v_{ts}^2} \right]
\label{eqn:F0.fpc}
\end{equation}
where $v_{dsj}$ is the population's drift velocity. The linear normal
mode behavior for such equilibria is governed by solutions of the
dispersion relation
\begin{equation}
\underline{\underline{D}}\left(\omega,k\right)=k^2\lambda_{De}^2 +
\sum_j \left(\frac{q_s}{q_e}\right)^2\frac{n_{sj}}{n_i}\frac{T_e}{T_s}
\left[1+ \xi_{sj} Z_0(\xi_{sj}) \right]
\label{eqn:disp.fpc}
\end{equation}
where $\underline{\underline{D}}$ is a function of wavenumber $k$ and
complex frequency $(\omega,\gamma)$, $q_s$ is the species charge,
$Z_0$ is the plasma dispersion function\cite{Fried:1961} with argument
$\xi_{sj}=\left(\omega/ \omega_{pe} k \lambda_{De} \right)
\left(\sqrt{T_e m_s/2 T_s m_e} \right) - v_{dsj}/v_{ts}$, with the sum
taken over the three plasma populations $j$. The electron plasma
frequency $\omega_{pe} \equiv \sqrt{4 \pi \sum_j n_{ej}q^2/m_e}$ and
Debye length $\lambda_{De}=\sqrt{T_e/4 \pi \sum_jn_{ej}q^2}$, both
defined using the total electron density, normalize the time and
length scales of our system. Values for the normalized density of
electron population $j$, $n_{ej}/n_i$, and bulk velocity, normalized
by the electron thermal velocity $v_{te}=\sqrt{T_e/m_e}$, are given in
Table~\ref{tb:params}. For all three cases, the electron populations
have equal temperatures; the ions are singly ionized, and initialized
with $T_i=T_e$, $m_i=100m_e$, and $v_{di}=0$. Case 1 is stable to the
effects of the counter-streaming electrons, while the increase in
$|v_{dej}|$ for case 2 yields the classic counter-streaming
instability. Case 3 is unstable to the bump-on-tail instability.
Cases 2 and 3 serve as examples of fluid and kinetic instabilities
respectively.

In Fig~\ref{fig:linear}, solutions to Eqn~\ref{eqn:disp.fpc} for the
three cases are presented. In panel a, the complex frequency solutions
$(\omega, \gamma)/\omega_{pe}$ are given for fixed
$k\lambda_{De}=0.2$. For case 1 (black circles) all modes are shown to
be damped. For case 2 (red triangles), the increase in $|v_{dej}|$
leaves the frequency and damping rate of the Langmuir modes, the modes
with $|\omega| \approx \omega_{pe}$, largely unaffected. The pair of
least damped acoustic modes from case 1 are now both unstable, having
$\omega=0$ and two distinct growth rates, $\gamma>0$. The parametric
path from stability ($|v_{dej}|=0.0$, open triangles) to instability
($|v_{dej}|=1.75$, filled triangles) for these two modes is
illustrated as a function of complex frequency in panel b. For case 3,
the $\omega<0$ Langmuir mode and the least damped acoustic modes are
negligibly affected by the bump distribution. By parametric variation
of $v_{de2}$ from $0.0$ (open diamonds in panel c) to $4.25$ (filled
diamonds), it is observed that the growing mode for the bump-on-tail
instability arises from an acoustic mode which is strongly damped in
the stable regime (solid line), while the $\omega>0$ Langmuir mode
(dashed line) becomes heavily damped. The dispersion relations for the
least damped and/or fastest growing Langmuir (acoustic) modes as a
function of $k \lambda_{De}$ for the three cases are presented in
panels d and f (e and g) illustrating the wavelengths for which linear
instabilities arise for cases 2 and 3 (solid lines in panel g).

\section{Numerical Simulations}
\label{sec:vp}
To evaluate the time evolution of these systems, we have extended the
Vlasov-Poisson solver \texttt{VP}\cite{Howes:2016} to allow for the
inclusion of an arbitrary number of drifting plasma populations.
\texttt{VP} solves the nonlinear Vlasov-Poison system using
second-order finite differencing for spatial and velocity derivatives
and a third-order Adams-Bashforth scheme in time. As a test of this
extension, we compare numerical solutions of Eqn.~\ref{eqn:disp.fpc}
for the three cases described in section \ref{sec:drifts}
to frequencies and damping rates extracted from time traces of the
electrostatic field energy from linear and small-amplitude nonlinear
\texttt{VP} simulations for a range of wavevectors, given as points in
Fig.~\ref{fig:linear}. Agreement between the dispersion relation and
\texttt{VP} are close as long as the mode of interest is not heavily
damped and unstable modes supported by the system do not grow too
quickly with respect to the damped modes.

\begin{figure*}[t]
\hspace*{0.cm}
\includegraphics[width=16.5cm,viewport=0 5 550 135, clip=true]
{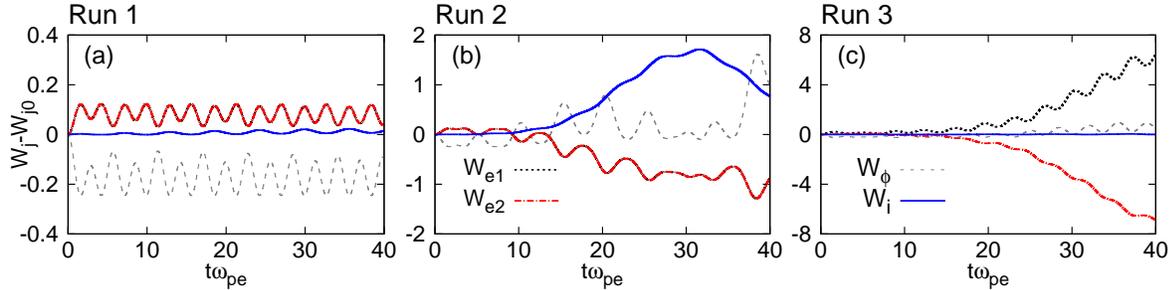}
\caption{The change in energy, $W_j(t)-W_j(t=0)$, for the two
  populations of electrons, the ions, and the electrostatic field
  (black, red, blue, and grey lines), for the three nonlinear
  simulations, normalized by $T_e$.}
\label{fig:energy}
\end{figure*}

For the evaluation of the field-particle correlation, we perform three
nonlinear simulations corresponding to the three cases from section
\ref{sec:drifts}. For these simulations, we add a sinusoidal
perturbation to the ion's equilibrium distribution of the form $0.1
F_{i0}(x,v) \sin(k_0 x)$ with $k_0\lambda_{de}=0.2$. 256 (128) points
in velocity (co\"ordinate) space are resolved over the interval
$v/v_{ts}\in[-8,8]$ $(x /\lambda_{De}\in[-5 \pi,5 \pi])$. Each
simulation is run for longer than $t = 40 \omega_{pe}^{-1}$ and the
total energy in the system
\begin{equation}
W_{\rm total}=\int dx \frac{E^2}{8 \pi} + \sum_s \int dx \int dv
\frac{m_s v^2}{2} f_s
\label{eqn:energy.fpc}
\end{equation}
is conserved to better than a few tenths of a percent.  Changes in the
electrostatic energy $W_\phi = \int dx E^2/8 \pi$ as well as energy in
the three plasma populations, $W_j = \int dx \int dv m_j v^2 f_j/2$,
from their initial values are shown in Fig.~\ref{fig:energy}, with
energy normalized by $T_e$. In run 1, most of the energy damped from
the electric field is equally partitioned between the two electron
populations, with little energy transferred to the ions. For run 2, an
instability is clearly triggered, with significant losses of energy
from the electrons and both $W_\phi$ and $W_i$ growing from their
initial values. A more virulent instability is triggered in run 3,
with the bump electron population losing a significant fraction of its
energy to the core electron population, with little energy transferred
to the ions.

\section{Field-Particle Correlations}
\label{sec:fpc}
While tracking the change in energy is sufficient to identify
instabilities in systems with complete knowledge of spatial and
velocity structure, we seek the signature of unstable behavior given
limited, single-point measurements of a system of the type available
to spacecraft in the solar wind. The application of field-particle
correlations to such measurements allows for a local observation of
secular energy transfer.

We define the field-particle correlation for a discrete set of
measurements of $f_s(x,v,t)$ and $E(x,t)$ with timestep $dt$, taken at
a single point $x=x_0$ as
\begin{equation}
C_E(x_0,v,t_i,N)=-\frac{1}{N}\sum_{j=i}^{i+N}
\frac{q_sv^2}{2}\frac{\partial f_s(x_0,v,t_j)}{\partial v} E(x_0,t_j).
\label{eqn:FPC.fpc}
\end{equation}
The correlation averages the field-particle interaction term in the
Vlasov equation, the third term in Eqn.~\ref{eqn:vlasov.ld}, over a
time interval of length $\tau = N dt$. As the ballistic term, the
second in Eqn.~\ref{eqn:vlasov.ld}, does not lead to net energy
transfer,\cite{Howes:2016} the product in $C_E$ represents the energy
density transferred at one point in phase space between the electric
field and velocity distribution. By averaging over a selected time
interval $\tau$, oscillatory energy transfer between $E$ and $f_s$ is
removed, leaving only the secular energy transfer. To track the
accumulated change of the phase-space energy density, we integrate the
correlation over time, defining $\Delta w_s(x,v,t,N)\equiv \int_0^t
dt' C_E(x,v,t',N)$.

\begin{figure}[t]
\hspace*{0.cm}
\includegraphics[width=16.5cm,viewport=5 25 550 135, clip=true]
{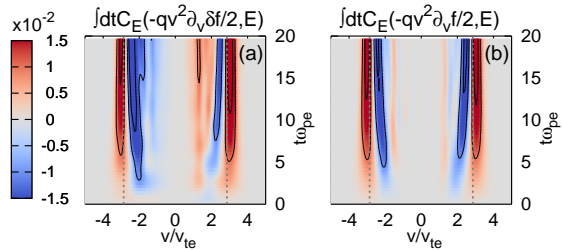}
\caption{The accumulated change in the electron phase-space energy
  density for damped, counter-propagating Langmuir waves calculated
  using the perturbed electron distribution in panel a, Eqn.~2 from
  Klein \& Howes 2016\cite{Klein:2016a} and the total electron
  distribution in panel b, Eqn.~\ref{eqn:FPC.fpc}. The resonant
  velocities for the system are indicated by dashed grey lines.}
\label{fig:compare}
\end{figure}

Unlike Eqn.~2 in Klein \& Howes 2016,\cite{Klein:2016a} we include the
full distribution function in our definition of $C_E$, rather then
only the perturbed component $\delta f_s$. Previous studies had
focused on particular cases where the equalibria $F_{s0}$ were even
with respect to $v=0$, ensuring that they would not contribute to net
energy transfer. For the cases under consideration in this work, the
equilibrium electron distributions have odd components and therefore
may contribute to a secular transfer of energy between the fields and
distributions. To show that the two forms of the correlation obtain
similar results when $F_{s0}$ is even, we apply both correlations to
single-point field and distribution data from a Landau damped
counter-propagating Langmuir wave simulation, case 1 in Klein \& Howes
2016, and plot $\Delta w_e$ at $x=0$ with $\tau \omega_{pe} = 6.28$ in
Fig.~\ref{fig:compare}. We see that both correlations produce
qualitatively similar structure in the accumulated phase-space energy
density, especially in regards to production of a plateau surrounding
the resonant velocities of the system, $|v_{\rm res}|= 2.86 v_{te}$,
which serves as the key velocity-space signature of Landau damping.

\begin{figure*}[t]
\hspace*{0.cm}
\includegraphics[width=15.5cm,viewport=10 5 555 300, clip=true]
{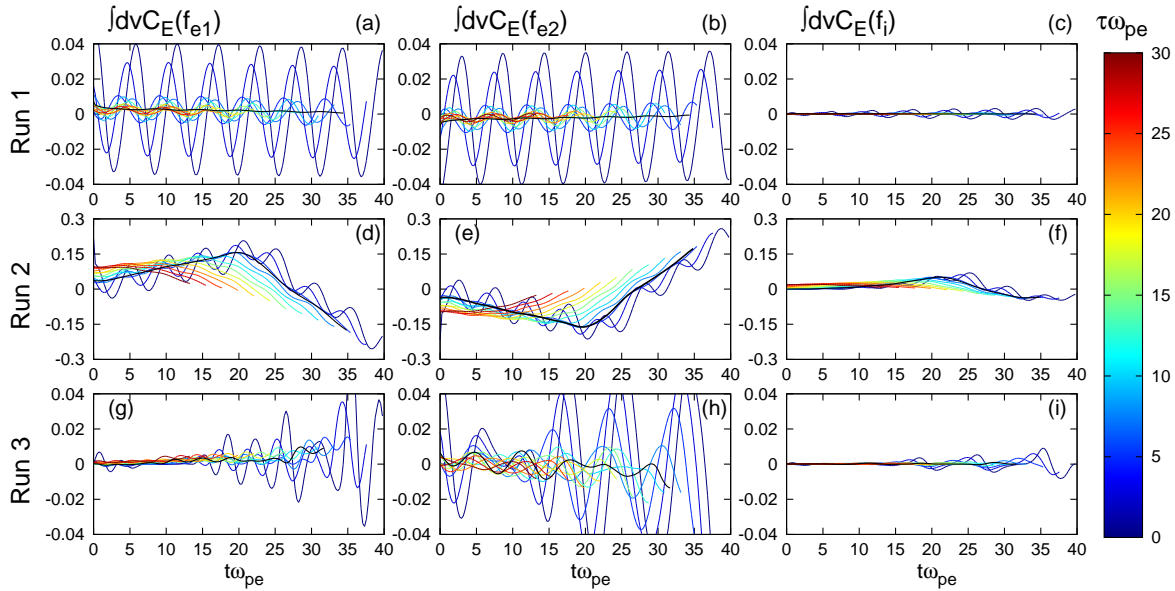}
\caption{The velocity integrated field-particle correlation $C_E(f_s)$
  at $x=0$ for a range of correlation intervals $\tau \in [0,30]$,
  differentiated in color. Each row represents the three simulations
  under consideration, while the columns present the correlation with
  $f_{e1}$, $f_{e2}$ and $f_i$. The black line highlights the $\tau$
  selected for Fig~\ref{fig:C_single}.}
\label{fig:tau}
\end{figure*}

With our correlation defined, we next select an appropriate
correlation interval $\tau$ for the three simulations.  By averaging
over a particular interval, we remove the transfer of energy between
the fields and particles which oscillates with frequency of the order
$\omega \sim 2 \pi/ \tau$, leaving the secular, or non-oscillatory,
component.  Plotted in Fig.~\ref{fig:tau} are velocity integrated
correlations, $\int dv C_E(x,v,t,\tau)$, between $E$ and the three
plasma populations for a range of correlation lengths $\tau
\omega_{pe} \in [0,30]$ at a single spatial location $x=0$.

There is significant oscillatory transfer for small $\tau$
correlations for run 1, panels a-c of Fig.~\ref{fig:tau}. This
oscillatory transfer is reduced for longer intervals, with nearly all
of the oscillations removed for $\tau \omega_{pe} = 5.64$, a
correlation length corresponding to the period of Langmuir waves
supported by the system with frequency $\omega = 1.11
\omega_{pe}$. This $\tau$ leaves the velocity integrated correlations
for all three distribution functions nearly monotonic, while slightly
longer correlations reintroduce some oscillatory behavior.  As the
Langmuir wave is the least damped, finite frequency linear mode
supported by this system, correlating over its period is physically
justified. For run 2, panels d-f, correlating over the interval $\tau
\omega_{pe} = 5.15$, which corresponds to the Langmuir wave frequency
$\omega = 1.21 \omega_{pe}$, removes significant oscillatory energy
transfer. Averaging over the time scale associated with the least
damped modes, rather than the unstable modes, is motivated by the fact
that the unstable modes of this system have zero frequency; see
Fig~\ref{fig:linear}. For run 3, we see that there does not exist a
single value of $\tau$ for which the oscillatory energy transfer is
completely removed. This due to the fact that the system supports both
a weakly damped Langmuir wave as well as a finite-frequency unstable
acoustic mode. Correlating over the Langmuir period retains some of
the acoustic oscillations, while correlating over the acoustic period
retains some of the Langmuir oscillations. We choose $\tau
\omega_{pe}=7.11$, corresponding to the period of the growing acoustic
mode with $\omega = 0.88 \omega_{pe}$ and acknowledge that some
oscillatory contributions from Langmuir waves will persist.

\section{Velocity-Space Structure of Energy Transfer}
\label{sec:method}
With an appropriate interval $\tau$ selected, we calculate the
velocity dependent field-particle correlation for the three
simulations. By retaining the velocity dependence, we are able to
address the question of where the energy transfer occurs in phase
space, and use the structure of this energy transfer to characterize
the nature of the underlying instability.

\begin{figure*}[t]
\hspace*{0.cm}
\includegraphics[width=15.5cm,viewport=5 20 395 345, clip=true]
{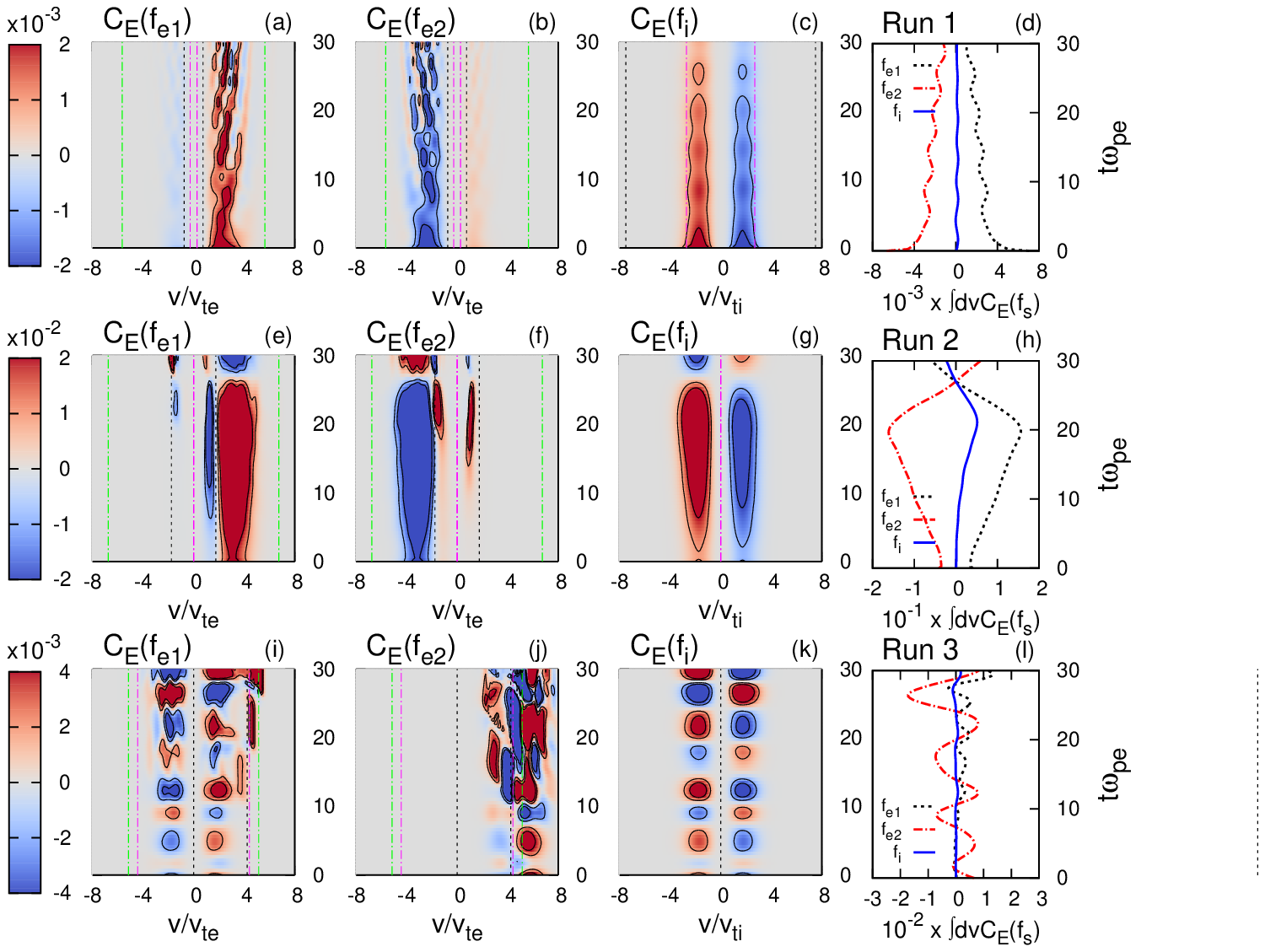}
\caption{The velocity dependent field-particle correlation $C_E(f_s)$
  at $x=0$ for runs 1 (panels a-c), 2 (panels e-g), and 3 (panels i-k)
  as well as the velocity integrated correlation for $f_{e1}$,
  $f_{e2}$, and $f_i$ (black, red, and blue lines in panels d, h, and
  l), for $\tau \omega_{pe} = 5.64$ (run 1), $5.15$ (run 2), and
  $7.11$ (run 3). The vertical dashed black lines indicate the
  electron drift velocities $v_{d1}/v_{te}$ and $v_{d2}/v_{te}$, while
  the green and magenta vertical lines give the resonant velocities
  for the least damped and/or fasted growing Langmuir and acoustic
  waves respectively.}
\label{fig:C_single}
\end{figure*}


For the stable, counter-streaming electron case, run 1, the
phase-space energy transfer obtained from the field-particle
correlation is a fairly regular function of velocity. At a given point
in co\"ordinate space, for example $x = 0 \lambda_{De}$ shown in the
first row of Fig.~\ref{fig:C_single}, one of the electron populations
gains energy from, while the other population loses energy to, the electric
field. The energy transfer to the ions from the electric field is an
odd function of velocity, meaning that when the correlation is
integrated over $v$, there is no net energy transfer to the
ions. There is no evidence for the dependence of the phase-space
energy transfer on drift velocities (black vertical dashed lines) or
the resonant velocities of either the Langmuir (green lines) or
acoustic (magenta) modes for any of the three populations.


Increasing the speed of the counter-streaming electrons for run 2,
shown in panels e-h of Fig.~\ref{fig:C_single}, the structure of the
energy transfer is altered. The sign of the field-particle correlation
changes at $v_{dej}$ due to a change in sign of $\p f_{ej}/\p v$. Unlike
for case 1, this change in sign occurs for electrons which exchange
significant energy with the electric field. For the ions, a small,
even component in the field-particle correlation arises, yielding a
net transfer of energy from the fields to the ions, as seen in panel
h. This transfer of energy to the ions, which does not depend on
either the Langmuir or acoustic resonant velocities, serves as a
phase-space signature for the fluid instability that arises for this
system.


For the bump-on-tail instability, case 3 shown in panels i-l of
Fig.~\ref{fig:C_single}, the velocity-space structure of the
field-particle correlation is significantly different. As we have
correlated over the unstable acoustic period, the oscillatory
structure from the weakly damped Langmuir mode is evident. We also see
in the bump electron population that the energy transfer changes sign
across the acoustic resonance and at later time across the Langmuir
resonance. This resonant structure is the signature of the transfer of
energy from the bump population to the core electron population as
mediated by the electric field. We note that the resonant structure is
maintained for other choices of $\tau$, including an interval
corresponding to the Langmuir wave period, not shown. The
velocity-integrated correlation confirms that the core population has
a net gain of energy, and as expected for this instability, we see
that the electrons that receive the energy are near the resonant
velocities. This explicit dependence of the phase-space energy
transfer on resonant coupling between the fields and particles serves
as a distinct signature of kinetic instabilities when compared to the
fluid instability in run 2.

\begin{figure*}[th]
\hspace*{0.cm} \includegraphics[width=15.5cm,viewport=0 5 410 210,
  clip=true] 
        {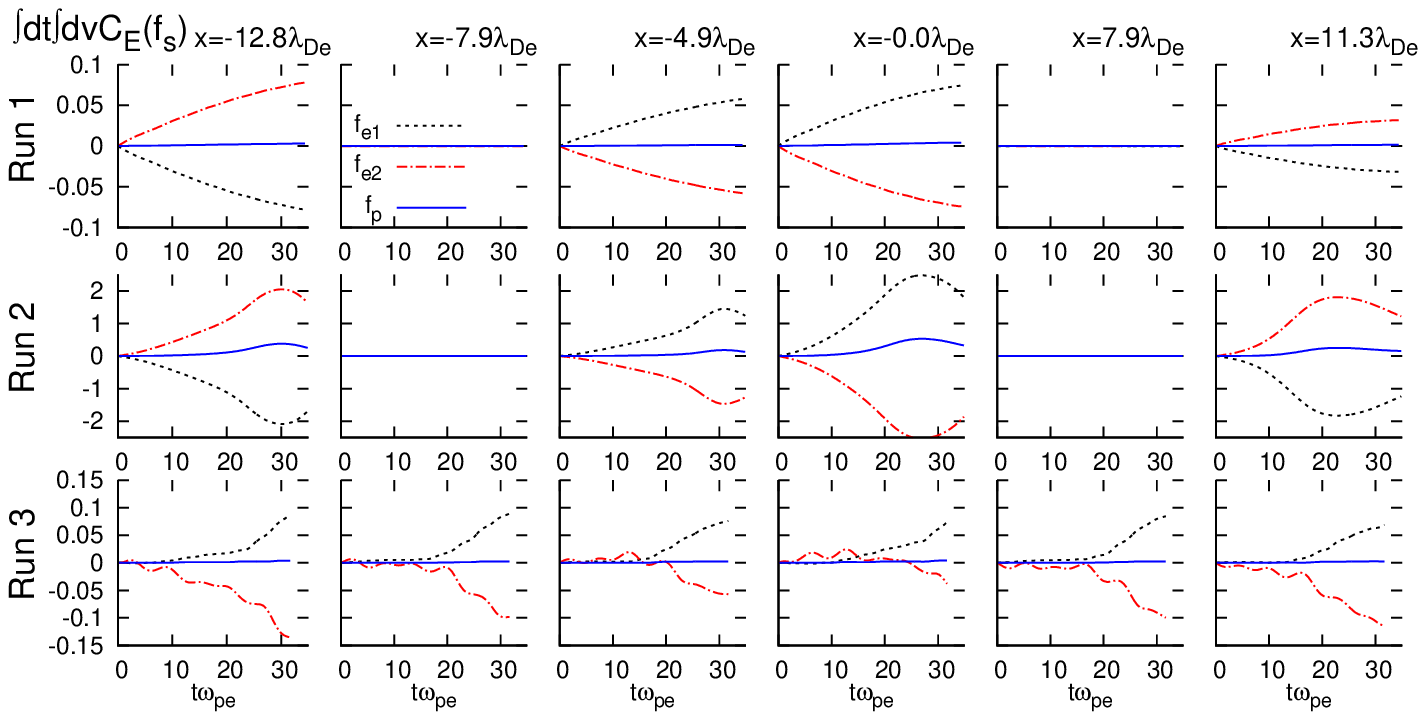}
\caption{The time and velocity integrated field-particle correlation
  $\int dt \int dv C_E(f_s)$ at six points in co\"ordinate space,
  arranged by columns, with values for $f_{e1}$, $f_{e2}$, and $f_i$
  given in black, red, and blue. The top, central, and bottom rows
  illustrate values for runs 1,2, and 3.}
\label{fig:int_C}
\end{figure*}

The field-particle correlations presented in Fig.~\ref{fig:C_single}
were calculated at a single point in co\"ordinate space. An obvious
question arises as to how the correlation and the associated
phase-space energy transfer change as a function of position within
the simulation. To asses this question, we calculate $C_E$ at five
other points within the simulation and present the velocity integrated
accumulated change of the phase-space energy density, $\int dv \Delta
w_s$, in Fig.~\ref{fig:int_C}. For the stable case, run 1, we see that
the energy transfer to or from the two electron populations changes
sign and amplitude as a function of the position in the simulation,
with the transfer passing through zero at the two nodes of the initial
standing wave pattern at $\pm 7.9 \lambda_{De}$. The correlation with
the ions maintains an even velocity space structure such that the ions
continue to neither lose nor gain net energy from the electric field
regardless of spatial position. For the unstable counter-streaming
electrons, run 2, the same pattern of shifting sign and amplitude for
the energy transfer to and from the electron holds. The ions gain
energy from the electric field regardless of which electron population
is gaining or losing energy.  The amplitude of the ion energy gain
changes with the amplitude of the electron field-particle correlation,
going to zero at the nodes and having its largest value at the
anti-nodes. This phase space structure serves as a distinct signature
of growing fluid instabilities. For the bump-on-tail instability, run
3, there is no regular spatial variation in the transfer of energy
between the beam and core electron populations; the core gains energy
at the expense of the beam, with the energy density accumulating at
nearly the same rate regardless of spatial position, as expected for a
kinetic instability.

\section{Conclusion}
\label{sec:conc}
Field-particle correlations of the form defined in
Eqn.~\ref{eqn:FPC.fpc}, modified from Klein \& Howes
2016\cite{Klein:2016a} to account for drifting equilibrium
distributions, are applied to a set of simulations where fluid and
kinetic instabilities are present. The structure of the resulting
correlations, which can be interpreted as the secular phase-space
energy density transferred between the fields and distributions, can
be used to identify the presence of instabilities as well as to
characterize the mechanisms driving the unstable growth. The form of
the correlation allows for such characterizations to be made from
observations at a single point, or a few points, in co\"ordinate
space, as opposed to requiring knowledge of spatially integrated
quantities typically not accessible to experimental measurements.

We consider simplified 1D-1V electrostatic systems in an attempt to
characterize field-particle correlations as measurements of secular
energy transfer in advance of future work applying such correlations
to systems of higher dimensionality, where magnetization, turbulence,
and inhomogeneities may complicate the interpretation of the
correlation. By determining signatures of basic plasma physics
phenomena responsible for energy transfer between fields and
distributions, such as Landau damping and one-dimensional
instabilities, we lay the foundation for the determination of the
velocity distribution signatures of more complex interactions, such as
cyclotron\cite{Stix:1992} and transit time damping,\cite{Barnes:1966}
heating by large amplitude, stochastic
fluctuations,\cite{Chandran:2010a} and magnetic
reconnection,\cite{Yamada:2010} which have all been proposed to play a
role in the dissipation of turbulent fluctuations. Future work will
also consider the effects of solar wind expansion, electron
conduction, and other inhomogeneous mechanisms on the plasma to
clearly identify the role of such energy transfer mechanisms in the solar wind. By
constructing the correlation to be obtained from single-point
measurements, we allow for the identification of such damping
mechanisms from in situ observation of the solar wind on current and
future missions including \textit{Deep Space Climate Observatory
(DSCOVR)}, \textit{MMS},\cite{Burch:2016} and \textit{Solar Probe
Plus}.\cite{Fox:2015}

\begin{acknowledgments}
The author would like to thank Gregory Howes, Justin Kasper, and Jason
TenBarge for insightful discussions regarding aspects of this work.
This research was supported by the NASA HSR grant NNX16AM23G.
\end{acknowledgments}

\bibliographystyle{aip}


\end{document}